\documentclass[aps,pra,reprint,groupedaddress]{revtex4-2}

\usepackage{amsmath}
\usepackage{graphicx}
\usepackage{dcolumn}
\usepackage{bm}
\usepackage[colorlinks,linkcolor=blue,anchorcolor=blue,citecolor=blue,urlcolor=blue]%
{hyperref}
\usepackage{epsfig}

\begin{document}

\title{Mechanical Squeezing via Detuning-Switched Driving}

\author{Yaohua Li$^{1}$}
\author{An-Ning Xu$^{1}$}
\author{Long-Gang Huang$^{1}$}
\author{Yong-Chun Liu$^{1, 2}$}
\email{ycliu@tsinghua.edu.cn}
\affiliation{$^{1}$State Key Laboratory of Low-Dimensional Quantum Physics, Department of Physics, Tsinghua University, Beijing 100084, P. R. China}
\affiliation{$^{2}$Frontier Science Center for Quantum Information, Beijing 100084, P. R. China}

\date{\today}

\begin{abstract}
	Generation of mechanical squeezing has attracted a lot of interest for its nonclassical properties, applications in quantum information, and high-sensitivity measurement. Here we propose a detuning-switched method that can rapidly generate strong and stationary mechanical squeezing. The pulsed driving can dynamically transpose the optomechanical coupling into a linear optical force and maintain an effective mechanical frequency, which can introduce strong mechanical squeezing in a short time. Moreover, we show the obtained strong mechanical squeezing can be frozen by increasing the pulse intervals, leading to stationary mechanical squeezing with a fixed squeezing angle. Thus, our proposal provides fascinating insights and applications of modulated optomechanical systems.
\end{abstract}

\maketitle

\section{Introduction}

Mechanical squeezed states allow the quantum fluctuation of a single quadrature below the zero-point fluctuation. The unavoidable fluctuations limit the precision of measurement of mechanical quadratures, which then can be surpassed with the mechanical squeezed states \cite{braginsky_quantum_1980,burd_quantum_2019}. Mechanical squeezing was firstly considered and realized in the parametric resonators \cite{PhysRevA.36.2463,PhysRevLett.67.699}, the frequency of which was modulated at twice the original mechanical oscillation frequency. In optomechanical systems, the parametric squeezing can be realized by the modulation of the optical spring \cite{ISI:000168559600048,PhysRevA.83.033820,PhysRevLett.112.023601,PhysRevA.96.063811,PhysRevA.98.013804}. However, the steady-state squeezing degree is limited to $3\mathrm{dB}$ due to the divergence of the amplified quadrature \cite{PhysRevLett.67.699}. To obtain strong mechanical squeezing, additional measurement and appropriate feedback control are required to suppress the amplification \cite{PhysRevLett.107.213603,PhysRevLett.110.184301,PhysRevLett.111.207203,PhysRevA.98.013804}. Conversely, measurements can also be used to prepare quantum states, especially the mechanical squeezed states \cite{Clerk_2008,ISI:000275024000023,PhysRevX.5.041037,ISI:000360646800047}. With sufficiently strong measurements and optimal estimation of the quadrature, strong mechanical squeezing can be obtained in a conditional state \cite{PhysRevLett.123.093602,PhysRevLett.125.043604}. However, feedback force related to the estimation results is also required to convert the system into unconditional squeezing \cite{PhysRevA.60.2700,PhysRevA.88.063833,PhysRevLett.115.243601}.

Instead of a long dissipative evolution, strong mechanical squeezing can be also obtained in nonequilibrium processes, for example, via the rapid \cite{fan_squeezing_1988,PhysRevLett.67.3665,PhysRevLett.117.273601,xin_rapid_2021}, periodic \cite{PhysRevLett.103.213603,PhysRevA.100.023843,ISI:000511839000007} or pulsed \cite{PhysRevB.71.235407,ISI:000295255300018} modulations of the optical driving, and through the unstable multimode dynamics \cite{kustura_mechanical_2022}. Different designs of optical driving structures can touch different goals, including ultra-precise measurements and state preparation, without other assistance or additional feedback. State preparation with fast pulses has made great progress in the experiment \cite{ncomms3295,PhysRevLett.123.113601}. However, the amplitude-modulated structure requires the cavity decay rate much larger than the mechanical frequency to keep the pulse durations small after inputting the cavity \cite{ISI:000317589100002,ISI:000450211600003}, which precludes further squeezing of the mechanical mode. Moreover, the mechanical squeezing obtained from nonequilibrium processes is far from a steady state, and the rapid preparation of stationary mechanical squeezing remains a challenge.

\begin{figure}[t]
	\includegraphics[width=0.48\textwidth]{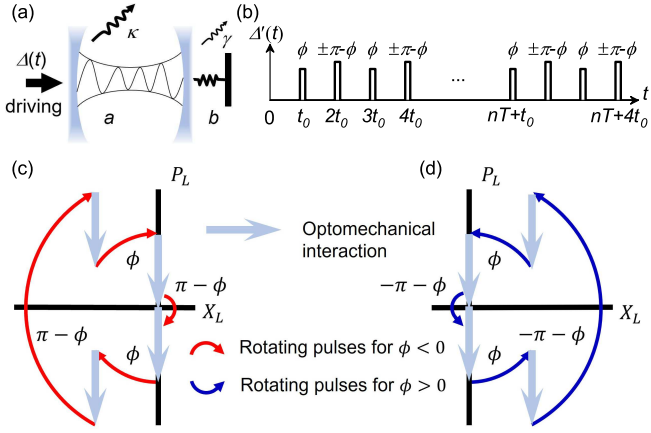}%
	\caption{Optomechanical model with detuned-switched driving. (a) Schematic of the optomechanical model in this work. The cavity mode is driven by a detuning-switched laser. (b) The detuning structure as a function of time with $\phi\in[-\pi,\pi]$. A series of four optical rotating pulses are applied to the optical mode through large detunings. (c),(d) Sketches of the evolution of optical quadrature operators in a four-pulse period for $\phi<0$ (c) and $\phi>0$ (d). The straight arrows and curved arrows indicate the evolutions between and during rotating pulses.\label{fig:scheme}}
\end{figure}

In this work, we firstly analyze the squeezing effect of a mechanical resonator induced by the detuning-switched driving in an optomechanical system [see Fig. \ref{fig:scheme}]. A series of four optical rotating pulses, which introduce an additional small period to the optical mode, yield quick measurements of mechanical position and induce a linear optical force. We can directly control the effective mechanical frequency and obtain pure squeezing terms through the optical force. Without additional readout and feedback control, the mechanical mode evolves to a deterministic squeezed state in a short time under the pulsed driving. By increasing the pulse interval, we further increase or decrease the effective mechanical frequency. In this way, the squeezed state becomes a thermal coherent state of the mechanical resonator under effective frequency, and we obtain stationary mechanical squeezing without a long dissipative evolution. Stationary mechanical squeezing allows improved measurement precision of a fixed quadrature, which is not available with conditional or rotated squeezing states.

\section{System model}

In the rotating frame with the frequency of the driving laser, such an optomechanical model can be well described by the Hamiltonian as $H=-\Delta(t) a^{\dag}a+\omega_{\mathrm{m}}b^{\dag}b+ga^{\dag}a(b+b^{\dag})+\Omega a^{\dag}+\Omega^{*} a$, where $a$ $(a^{\dag})$ and $b$ $(b^{\dag})$ are the optical and mechanical annihilation (creation) operator. $\omega_{\mathrm{m}}$ is the mechanical resonance frequency. $g$ is the single-photon optomechanical coupling strength. $\Omega$ describes the laser driving strength and $\Delta(t)$ is the time-dependent detuning between the driving laser and cavity resonance frequency. The optical (mechanical) decay rate is $\kappa$ ($\gamma$). Most of the time, $\Delta(t)=\Delta_{0}$ keeps a small value (compared with the enhanced optomechanical coupling strength G) except during the rotating pulse when the detuning is shifted to a huge value to rapidly change the relative phase of the cavity field. As the rotating pulse is very short (compared with the pulse interval), we can still employ the linearized process, i.e., $a\to a+\alpha$ and $b\to b+\beta$ where $\alpha=\langle a\rangle$ and $\beta=\langle b\rangle$ are the average amplitudes of the optical and mechanical mode. Then the linearized Hamiltonian can be obtained as (see Appendix \ref{ap:A} for more details)
\begin{equation}\label{eq:hl}
	H_{L}=-\Delta^{\prime}(t) a^{\dag}a+\omega_{\mathrm{m}}b^{\dag}b+(Ga^{\dag}+G^{*}a)(b+b^{\dag}),
\end{equation}
where $\Delta^{\prime}(t)=\Delta(t)-g(\beta+\beta^{*})$ is the effective detuning and $G=g\alpha$ is the enhanced optomechanical coupling strength.

The detuning during each pulse is so large that the evolution of the mechanical resonator and optomechanical coupling can be neglected. Then the evolution can be described by the rotating operators $R(\theta_{j})=e^{i\theta_{j} a^{\dag}a}$, where $\theta_{j}=\int_{jt_{0}}^{jt_{0}+\delta t}\Delta^{\prime}(\tau)d\tau$ is the rotating angle during each pulse and $\theta_{j}=\phi,\pm\pi-\phi,\phi,\pm\pi-\phi,\dots(\phi\in[-\pi,\pi])$ in our pulse structure [see Fig. \ref{fig:scheme}(b)]. As shown in Fig. \ref{fig:scheme}(c) and \ref{fig:scheme}(d), the first two evolutions between rotating pulses in a four-pulse period yield quick measurement and store the information in quadrature $P_{L}$ which then be erased and renewed soon within the rotating pulses [see Fig. \ref{fig:scheme}(c) and \ref{fig:scheme}(d)]. The rotating pulses also transpose the information to quadrature $X_{L}$ which can react on the mechanical quadratures soon in the later optomechanical coupling. The information transposed by pulses with rotating angles $\phi>0$ [see. Fig. \ref{fig:scheme}(c)] and $\phi<0$ [see. Fig. \ref{fig:scheme}(d)] is different, which results in opposite feedback controls of the mechanical quadratures. These two sketches are plotted according to the Heisenberg equations [see Eq. (\ref{eq:measure}) and (\ref{eq:m2}) in the following]. We note that the cavity quadratures will not rotate following the driving laser if we only employ a sudden switch of the laser frequency. It also requires a stronger power of the driving laser to compensate for the larger detuning, and the phase of the laser driving should also be adjusted. Moreover, the detuning should not be enlarged arbitrarily fast, but in  a time scale smaller than the round trip time of the cavity. In this case, the cavity can still reach a new equilibrium with the large detuning [see Appendix \ref{ap:dis} for more details].

It's convenient to employ the Baker-Campbell-Hausdorff formula here to analyze the whole evolution operator in a four-pulse period (the total rotating phase of the optical field in a period is $2\pi$), which can be written as
\begin{equation}
	\mathcal{U}(T)=\left[R(\pi-\phi)U(t_{0})R(\phi)U(t_{0})\right]^2,
\end{equation}
where $T=4t_{0}$ is the four-pulse period and $U(t_{0})=e^{-iH_{L}t_{0}}$ describes the evolution between two pulses. The effective detuning between two pulses is $\Delta_{0}-g(\beta+\beta^{*})\equiv\Delta_{0}^{\prime}$. Here we employ an approximation that the pulse interval is small, i.e., $\omega_{\mathrm{m}}t_{0}\ll 1$. Then the higher order terms in the BCH formula can be directly neglected except for the first and second ones, leading to a Hamiltonian that satisfies $\mathcal{U}(T)=e^{-iH_{\mathrm{0}}T}$. The Hamiltonian can be written as $H_{0}=H_{\mathrm{eff}}+H^{\prime}$, where
\begin{equation}\label{eq:heff}
	H_{\mathrm{eff}}=\omega_{\mathrm{m}}b^{\dag}b+\sigma(b+b^{\dag})^{2},
\end{equation}
\begin{equation}
	\sigma=\frac{1}{2}|G|^{2}t_{0}\mathrm{sin}\phi,
\end{equation}
and $H^{\prime}=-\Delta_{0}^{\prime} a^{\dag}a+\Delta_{0}^{\prime}(\mu a^{\dag}+\mu^{*}a)(b+b^{\dag})+\omega_{\mathrm{m}}(\mu a^{\dag}-\mu^{*}a)(b-b^{\dag})$, $\mu=\frac{1}{4}Gt_{0}(1+e^{i\phi})$  (see Appendix \ref{ap:B} for more details). In the strong coupling and small detuning regime that we consider, i.e., $|G\mathrm{sin}\phi|\gg |\Delta_{0}^{\prime}|,\omega_{\mathrm{m}}$, and with the assumption that $|\mu|\ll1$, the last two optomechanical coupling terms of $H^{\prime}$ can be neglected. Thus the first term of $H^{\prime}$ can be neglected together, leaving an effective Hamiltonian that contains pure mechanical squeezing terms $b^{2}+b^{\dag2}$ without any optomechanical couplings.

\section{Squeezing generation.}

To understand and quantify the obtained mechanical squeezing, we further introduce dimensionless quadratures defined as $X_{L}=a+a^{\dag}$, $P_{L}=i(a^{\dag}-a)$, $X_{M}=b+b^{\dag}$, $P_{M}=i(b^{\dag}-b)$ and $Y_{M}=be^{-i\theta/2}+b^{\dag}e^{i\theta/2}$, where $Y_{M}$ is the squeezed mechanical quadrature and $\theta$ is the squeezing angle. So the Hamiltonian in Eq. (\ref{eq:heff}) can be rewritten as $H_{\mathrm{eff}}+1/2=\omega_{\mathrm{m}}(X_{M}^{2}+P_{M}^{2})/4+\sigma X_{M}^{2}$. It means that the optomechanical coupling can be dynamically transposed into a constant modulation of the potential energy or the spring constant felt by the mechanical resonator. In other words, such pulse structure can rapidly maintain an effective mechanical resonance frequency as
\begin{equation}\label{eq:os}
	\omega_{\mathrm{s}}=\sqrt{\omega_{\mathrm{m}}(\omega_{\mathrm{m}}+4\sigma)},
\end{equation}
in the dynamically stable regime for $\sigma>-\omega_{\mathrm{m}}/4$. $\sigma=-\omega_{\mathrm{m}}/4$ is the threshold condition of the pulsed driving. When above the threshold, the variance of amplified mechanical quadrature increases exponentially. Noted that the threshold only exists when the rotating angle $\phi<0$ when the effective potential energy of the mechanical resonator is reduced to zero.

It is more clear from the Heisenberg equations of quadratures. Without losing generality, we assume $G$ was real. Then the linear Hamiltonian in Eq. (\ref{eq:hl}) can be rewritten as
\begin{equation}
	H_{L}=-\frac{1}{4}\Delta^{\prime}(t)(X_{L}^{2}+P_{L}^{2})+\frac{1}{4}\omega_{m}(X_{M}^{2}+P_{M}^{2})+GX_{L}X_{M},
\end{equation}
and the Heisenberg equations are given by
\begin{gather}\label{eq:hh}
	\dot{X_{L}}=-\Delta^{\prime}(t) P_{L},\,\,\dot{P_{L}}=\Delta^{\prime}(t) X_{L}-2GX_{M},\\
	\dot{X_{M}}=\omega_{m}P_{M},\,\,\dot{P_{M}}=-\omega_{m}X_{M}-2GX_{L}.\label{eq:hh2}
\end{gather}
Between two pulses, we have $\Delta^{\prime}(t)=\Delta_{0}^{\prime}\ll G$. Eq. (\ref{eq:hh}) can be reduced to
\begin{gather}\label{eq:measure}
	\dot{X_{L}}\approx0,\,\,\dot{P_{L}}\approx-2GX_{M}.
\end{gather}
And during the pulses, $\Delta^{\prime}(t)\gg G$, we have
\begin{gather}\label{eq:m2}
	\dot{X_{L}}\approx-\Delta^{\prime}(t) P_{L},\,\,\dot{P_{L}}\approx\Delta^{\prime}(t) X_{L}.
\end{gather}
Fig. \ref{fig:scheme}(c), and \ref{fig:scheme}(d) are plotted according to Eq. (\ref{eq:measure}) and (\ref{eq:m2}). As shown in Eq. (\ref{eq:measure}), the optomechanical coupling between two nearest pulses generates a measurement on the mechanical quadrature $X_{M}$. Then the rotating pulse transposes the information into another optical quadrature $X_{L}$ and changes the optical force. The average optical force can be obtained as
\begin{equation}
	F_{\mathrm{ave}}=-\sigma X_{M}.
\end{equation}
which is proportional to the mechanical quadrature $X_{M}$. If $\phi=0,\pm\pi$, the average optical force is zero with no squeezing effect. There are similar quick measurements, but the information is only stored in $P_{L}$ and does not react on the mechanical quadratures. If $\phi=\pm\pi/2$, the information is transposed totally and maximal squeezing degree can be obtained.

\begin{figure}[t]
	\includegraphics[width=0.48\textwidth]{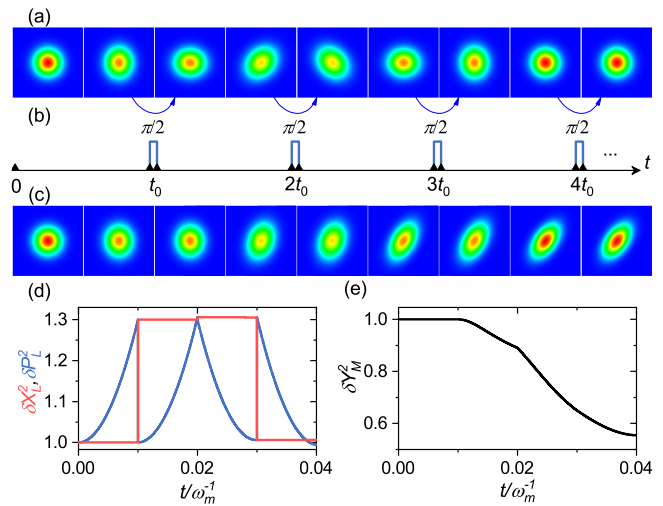}%
	\caption{Generation of mechanical squeezing. (a),(c) The Wigner functions of the optical (a) and mechanical (c) modes at certain times labeled in (b) by black triangles. The vertical axis is $X_{i}$ and the horizontal axis is $P_{i}$ for $i=L,M$. (b) The detailed detuning structure in the first four-pulse period, corresponding to $\phi=\pi/2$ in Fig. \ref{fig:scheme}(b). (d),(e) The time evolutions of optical (d) and mechanical (e) quadrature variances. The initial photon and phonon numbers are both zero. Other parameters are $\omega_{\mathrm{m}}t_{0}=0.01$, $\omega_{\mathrm{s}}=4\omega_{\mathrm{m}}$, $\Delta_{0}^{\prime}=0$,\label{fig:short}}.
\end{figure}

As an example for $\phi=\pi/2$, we plot the detailed Wigner functions and the evolution of quadrature variances in the first four-pulse period, starting with both optical and mechanical modes vacuum states [see Fig. \ref{fig:short} and Appendix \ref{ap:D}]. Without the pulses, the optomechanical coupling, as called X-X coupling, will generate large amplification of optical quadrature $P_{L}$ due to large optomechanical coupling strength. Additional pulses turn the trend around and, at the same time, transpose the information from $P_{L}$ to $X_{L}$. The latter can react to the mechanical quadratures, leading to a linear optical force and an effective mechanical frequency. Moreover, after the four-pulse sequence, all the mechanical information written on the optical field is erased, and the mechanical quadratures are correlated. It means that the mechanical squeezing obtained is very pure without measurement.

\begin{figure}[t]
	\includegraphics[width=0.48\textwidth]{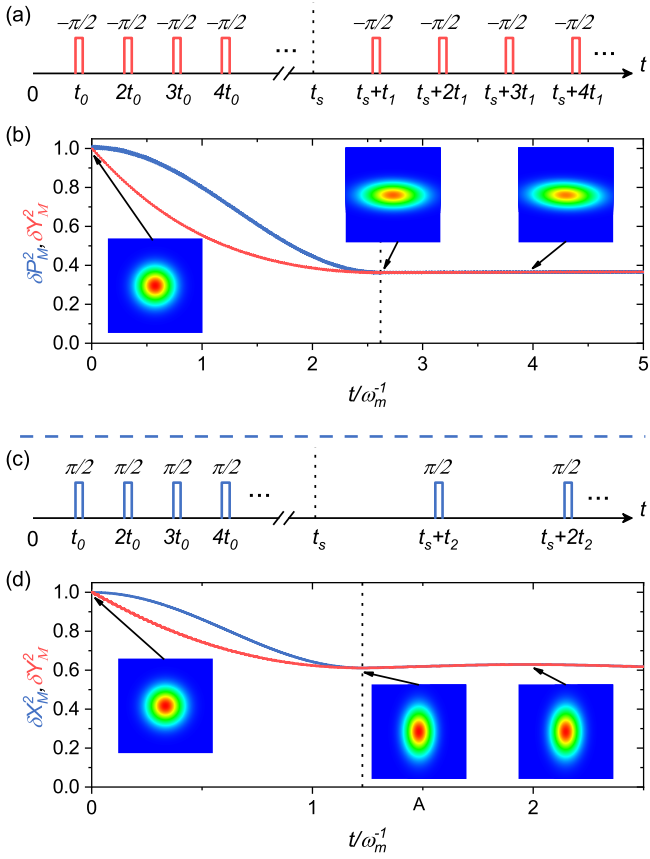}%
	\caption{Stationary mechanical squeezing. (a),(b) $\phi=-\pi/2$, (c),(d) $\phi=\pi/2$. (a),(c) The detuning structures to realize stationary mechanical squeezing. The pulse interval is increased to $t_{1,2}$ after obtaining maximum squeezing degree, i.e., $t=t_{\mathrm{s}}$. (b),(d) The time evolutions of mechanical quadrature variances for $\omega_{\mathrm{m}}t_{0}=0.005$ and $G/\omega_{\mathrm{m}}=8$. The insert figures are Wigner functions of mechanical mode at three certain times denoted by the arrows. Other unspecified parameters are the same as in Fig. \ref{fig:short}.\label{fig:min}}
\end{figure}

\section{Stationary mechanical squeezing}

As the pulsed driving rapidly maintains an effective mechanical frequency $\omega_{\mathrm{s}}$, the mechanical quadratures will rotate with the frequency, accompanied by quadrature squeezing with periodically changed squeezing degree and squeezing angle. The approximate time evolution of mechanical quadratures can be derived from the effective Hamiltonian $H_{\mathrm{eff}}$ and corresponding master equations. For $\sigma>-\omega_{\mathrm{m}}/4$ below the threshold, the evolution of variance of squeezed mechanical quadrature is given by (see Appendix \ref{ap:E} for more details)
\begin{equation}\label{eq:ymt}
	\begin{split}
		\delta Y_{M}^{2}&(t)=(1+2n_{\mathrm{th}})\\
		&\left[1-\frac{2\sigma^{2}}{\omega_{\mathrm{s}}^{2}}\cos^{2}\omega_{\mathrm{s}}t\left(\sqrt{1+\frac{\omega_{\mathrm{s}}^{2}}{\sigma^{2}\cos^{2}\omega_{\mathrm{s}}t}}-1\right)\right],
	\end{split}
\end{equation}
where $n_{\mathrm{th}}$ is the phonon number of the initial mechanical state. We have neglected the mechanical decay rate, as we are only interested in the short-time evolution far from equilibrium. The maximal squeezing degree is obtained at half the evolution period, i.e.,
\begin{equation}
	t_{\mathrm{s}}=\frac{\pi}{2\omega_{\mathrm{s}}}.
\end{equation}
And the minimal variances of squeezed quadrature are approximately given by
\begin{equation}\label{eq:rm}
	(\delta Y_{M}^{2})_{\mathrm{min}}=\left\{
	\begin{aligned}
		 & (1+2n_{\mathrm{th}})\frac{\omega_{m}+4\sigma}{\omega_{\mathrm{m}}},          & -\frac{\omega_{\mathrm{m}}}{4}<\sigma<0 \\
		 & (1+2n_{\mathrm{th}})\frac{\omega_{\mathrm{m}}}{\omega_{\mathrm{m}}+4\sigma}. & \sigma>0~~~~~
	\end{aligned}
	\right.
\end{equation}
The initial phonon number can be very low ($n_{\mathrm{{th}}}\lesssim1$) as we can employ laser cooling before the generation of mechanical squeezing. It means that our proposal can squeeze a mechanical mode with quadrature variance far below the zero-point fluctuation.

The maximally squeezed state at $t=t_{\mathrm{s}}$ is a thermal coherent state of the mechanical resonator with another frequency, $\omega_{\mathrm{s}}^{\prime}=\omega_{\mathrm{m}}+4\sigma$. The effective Hamiltonians with these two frequencies are given by
\begin{gather}
  H_{\mathrm{eff}}=\frac{1}{4}\omega_{\mathrm{s}}\left(\sqrt{\frac{\omega_{\mathrm{m}+4\sigma}}{\omega_{\mathrm{m}}}}X_{M}^{2}+\sqrt{\frac{\omega_{\mathrm{m}}}{\omega_{\mathrm{m}+4\sigma}}}P_{M}^{2}\right)-\frac{1}{2}\\
  H_{\mathrm{eff}}=\frac{1}{4}\omega_{\mathrm{s}}^{\prime}\left(X_{M}^{2}+\frac{\omega_{\mathrm{m}}}{\omega_{\mathrm{m}+4\sigma}}P_{M}^{2}\right)-\frac{1}{2}
\end{gather}
The mechanical quadratures with frequency $\omega_{\mathrm{s}}^{\prime}$ is unsqueezed at $t=t_{\mathrm{s}}$, i.e. $({\omega_{\mathrm{m}}+4\sigma})\delta X_{M}^{2}={\omega_{\mathrm{m}}}\delta P_{M}^{2}$. It means that the squeezed state is a thermal state with the effective frequency $\omega_{\mathrm{s}}^{\prime}$ and its Wigner function is a circle. The thermal state will be kept in the later evolution. Consequently, if we increase the pulse interval for $t>t_{\mathrm{s}}$ and thus change the effective mechanical frequency to $\omega_{\mathrm{s}}^{\prime}$, the mechanical mode will keep stationary at the maximally squeezed state, with both the squeezing degree and the squeezing angle unchanged. The idea of generating stationary squeezing by changing the frequency is firstly proposed by Ref. \cite{xin_fast_2022}. Here we can realize it by increasing the pulse interval. The increased pulse intervals are
\begin{equation}\label{eq:tp}
	t^{\prime}=2t_{0}\left[1+\frac{|G|^{2}t_{0}}{\omega_{\mathrm{m}}}\sin\phi\right].
\end{equation}
The detailed pulse structures are shown in Fig. \ref{fig:min}(a) and \ref{fig:min}(c), where $t_{1}$ ($t_{2}$) is the increased pulse interval for $\phi=-\pi/2$ ($\pi/2$). In Fig. \ref{fig:min}(b) and \ref{fig:min}(d), we plot the time evolutions of mechanical quadratures in the two cases. The mechanical mode evolves to the maximally squeezed state at $t=t_{\mathrm{s}}$. Afterward, we increase the pulse interval and obtain stationary mechanical squeezing in a short time. In particular, mechanical quadrature $P_{M}$ is squeezed ($Y_{M}=P_{M}$) with a squeezing angle $\theta=\pi$ for $\phi<0$, while for $\phi>0$ the squeezed mechanical quadrature is $Y_{M}=X_{M}$ with a squeezing angle $\theta=0$.

\begin{figure}[t]
	\includegraphics[width=0.47\textwidth]{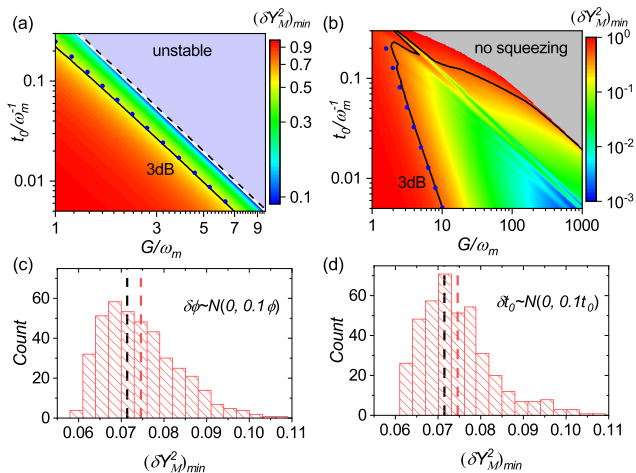}%
	\caption{Influence of parameters and Gaussian errors. (a),(b) The minimal variance of $Y_{M}$ as a function of $G$ and $t_{0}$ for $\phi=-\pi/2$ (a) and $\phi=\pi/2$ (b). The black solid line (blue dots line) indicates the contour line for 3dB squeezing of numerical (analytical) results. The light blue (grey) shadow indicates the unstable (no-squeezing) region. (c),(d) The histogram of minimal variances of $Y_{M}$ in 400 calculation events when adding Gaussian error to either rotating angles (c) or pulse intervals (d) of each pulse. The standard deviations of Gaussian errors are both one-tenth of the average values, which is $\phi=\pi/2$ and $t_{0}=0.01\omega_{\mathrm{m}}^{-1}$. The average results are $(\delta Y_{M}^{2})_{\mathrm{min}}=0.0746$ (red dashed line in panel (c)) and $(\delta Y_{M}^{2})_{\mathrm{min}}=0.0745$ [red dashed line in panel (d)]. Without the Gauss error, the minimal variance is $(\delta Y_{M}^{2})_{\mathrm{min}}=0.0714$ (black dashed line in panels (c) and (d)). Other parameters are $\Delta_{0}^{\prime}=0$, $\omega_{\mathrm{s}}=4\omega_{\mathrm{m}}$, i.e. $G\approx27.4\omega_{\mathrm{m}}$.\label{fig:para}}
\end{figure}

\section{Squeezing performance}

To verify the accuracy of the approximations made in the derivation of effective Hamiltonian, we plot the minimal variance of squeezed quadrature $Y_{M}$ as a function of $G$ and $t_{0}$ for $\phi=-\pi/2$ [see Fig. \ref{fig:para}(a)] and $\phi=\pi/2$ [see Fig. \ref{fig:para}(b)]. Large squeezing degrees can be obtained in a wide range of parameters, as predicted by Eq. (\ref{eq:rm}). But the approach works badly with parameters away from our assumptions ($\omega_{m}t_{0}\ll1$, $G\gg\omega_{m}$ and $|\mu|\ll1$, i.e., $|Gt_{0}|\ll1$). When $|Gt_{0}|\gtrsim1$, the optomechanical coupling terms that we used to neglect will greatly influence the evolution of mechanical mode and reduce the squeezing degree. For $\phi<0$, the parameter region $|Gt_{0}|\gtrsim1$ is deep into the unstable region which we are not interested in. To be clear, we emphasize the contour line of $3\mathrm{dB}$ squeezing (black solid lines), which agrees well with the exact values (blue dot lines) for appropriate parameters.

Furthermore, in Fig. \ref{fig:para}(c) and \ref{fig:para}(d), we analyze the influence of Gaussian errors that may exist in the parameters of pulse structure, i.e. $\phi$ and $t_{0}$. Typical Gaussian errors result in a variance of squeezing degrees with similar scaling compared with the average value. The average value of minimal variance is larger than the exact result without Gaussian error.

\section{Conclusion}

In conclusion, we propose a detuning-switched scheme that can dynamically generate strong and stationary mechanical squeezing in the optomechanical system rapidly. The detuning-switched pulses can transpose the optomechanical coupling into a linear optical force felt by the resonator and maintain an effective mechanical frequency, leading to a pure squeezing term. The squeezing process originates from a series of single-quadrature measurements and feedback through the optomechanical interactions, while the large-detuning pulses rotate the optical quadratures and allow the information of mechanical quadrature to be transferred, erased, and renewed in the optical quadratures. With this method, a large amount of squeezing can be achieved rapidly without a long dissipative evolution. Moreover, we show the squeezing degree and the squeezing angle at the maximally squeezed state can be frozen by increasing the pulse interval, i.e., further increasing or reducing the effective mechanical frequency. This method works well with large-range parameters and is robust to the Gaussian-shape error of pulse areas and pulse intervals. Our pulsed scheme can be also applied to other bosonic models that have a similar form of coupling, which provides a unique method to generate stationary squeezing in these systems.

\begin{acknowledgments}
	We are extremely grateful to the anonymous referees for their insightful comments on a better understanding of our scheme (e.g., discussions in Appendix F). This work is supported by the Key-Area Research and Development Program of
	Guangdong Province (Grant No.~2019B030330001), the National Natural Science
	Foundation of China (NSFC) (Grants No. 12275145, 92050110, 91736106, 11674390, and 91836302), and the National Key R\&D Program
	of China (Grants No. 2018YFA0306504).
\end{acknowledgments}

\appendix

\section{System Hamiltonian\label{ap:A}}

We consider a common optomechanical model that a Fabry-Perot cavity with cavity mode $a$ and frequency $\omega_{\mathrm{c}}$ is coupled with a mechanical oscillator with mechanical mode $b$ and frequency $\omega_{\mathrm{m}}$. To dynamically generate mechanical squeezing, a series of four optical rotating pulses with equal interval $t_{0}$ are included by rapidly changing the frequency of the driving laser, i.e. the detuning. The system Hamiltonian can be written as
\begin{equation}
	H_{\mathrm{S}}=\omega_{\mathrm{c}} a^{\dag}a+\omega_{\mathrm{m}}b^{\dag}b+ga^{\dag}a(b+b^{\dag})+(\Omega e^{-i\omega_{\mathrm{L}}(t)t}a^{\dag}+\mathrm{H.c.}),
\end{equation}
where $g$ is the single-photon optomechanical coupling strength and $\Omega$ [$\omega_{\mathrm{L}}(t)$] is the strength [frequency] of the optical driving. In the rotating frame with frequency $\omega_{\mathrm{L}}(t)$, the Hamiltonian can be rewritten as
\begin{equation}
	H=-\Delta(t) a^{\dag}a+\omega_{\mathrm{m}}b^{\dag}b+ga^{\dag}a(b+b^{\dag})+(\Omega a^{\dag}+\mathrm{H.c.}),
\end{equation}
where $\Delta(t)=\omega_{\mathrm{L}}(t)-\omega_{\mathrm{c}}$ is the laser detuning. We note here we have neglected the term containing the time derivative of the laser frequency, which is nonzero at the beginning and the end of the large-detuning pulses. Because this term will cancel each other at the beginning and the end of the large-detuning pulses as the pulse duration is extremely small. Between two pulses, the detuning $\Delta(t)=\Delta_{0}$ keeps a constant value. The quantum Langevin equations can be obtained as
\begin{gather}
	\dot{a}=[-\frac{\kappa}{2}+i\Delta(t)]a+iga(b^{\dag}+b)-i\Omega+a_{\mathrm{in}},\\
	\dot{b}=(-\frac{\gamma}{2}-i\omega_{\mathrm{m}})b+iga^{\dag}a+b_{\mathrm{in}},
\end{gather}
where $\kappa$ ($\gamma$) is the decay rate of the optical (mechanical) mode, and $a_{\mathrm{in}}$ and $b_{\mathrm{in}}$ are the corresponding noise operators. Here we employ the linearization process by replacing optical and mechanical operators with their average values and fluctuations, i.e. $a\to a+\alpha$ and $b\to b+\beta$, where the average values satisfy
\begin{gather}
	\dot{\alpha}=[-\frac{\kappa}{2}+i\Delta(t)]\alpha+ig\alpha(\beta^{*}+\beta)-i\Omega\label{eq:alpha},\\
	\dot{\beta}=(-\frac{\gamma}{2}-i\omega_{\mathrm{m}})\beta+ig|\alpha|^{2}\label{eq:beta}.
\end{gather}
Noted that $\alpha$ and $\beta$ given by Eq. (\ref{eq:alpha}) and (\ref{eq:beta}) are also time-dependent as the detuning is changed periodically. However, the time-dependent effect can be neglected as the pulse duration is small. And in principle, we can obtain constant solutions if the driving strength is also changed periodically. Then the quantum Langevin equations become
\begin{gather}
	\dot{a}=[-\frac{\kappa}{2}+i\Delta^{\prime}(t)]a+iga(b^{\dag}+b)+iG(b^{\dag}+b)+a_{\mathrm{in}}\label{eq:a},\\
	\dot{b}=(-\frac{\gamma}{2}-i\omega_{\mathrm{m}})b+iga^{\dag}a+i(Ga^{\dag}+G^{*}a)+b_{\mathrm{in}}\label{eq:b},
\end{gather}
where $\Delta^{\prime}(t)=\Delta(t)-g(\beta+\beta^{*})$ is the effective detuning and $G=g\alpha$ is the enhanced optomechanical coupling strength(We assume $G$ is a constant.). Considering small single-photon coupling strength and strong optical driving, the nonlinear terms $iga(b^{\dag}+b)$ and $iga^{\dag}a$ in Eq. (\ref{eq:a}) and (\ref{eq:b}) can be neglected and the linearized system Hamiltonian becomes
\begin{equation}\label{eq:linear}
	H_{L}=-\Delta^{\prime}(t) a^{\dag}a+\omega_{\mathrm{m}}b^{\dag}b+(Ga^{\dag}+G^{*}a)(b+b^{\dag}).
\end{equation}

\section{Baker-Campbell-Hausdorff formula and effective Hamiltonian\label{ap:B}}

In the Schr\"{o}dinger picture, the evolution of the optomechanical system is described by the operator $\mathcal{U}(t)=e^{-i\int H_{L}dt}$. Between two pulses, the effective detuning $\Delta^{\prime}(t)=\Delta_{0}-g(\beta^{*}+\beta)\equiv\Delta_{0}^{\prime}$ keeps a small and constant value. But during the rotating pulses, the effective detuning was rapidly enlarged to a huge value (in a time scale slower than the round trip time but much faster than t0). In this case, the evolution of mechanical mode as well as the optomechanical coupling, i.e. the second and third term in Eq. (\ref{eq:linear}) can be neglected. The system is well described by the rotating operator $R(\theta_{j})=e^{i\theta_{j} a^{\dag}a}$, where $\theta_{j}=\int_{jt_{0}}^{jt_{0}+\delta t} \Delta(\tau)d\tau$, $j=0,1,2,\cdots$ is the rotating angle during every pulse and $\theta_{j}=\phi,\pm\pi-\phi,\phi,\pm\pi-\phi,\cdots$ in our pulse structure. We define four pulses as a period because the total phase (angle) is $2\pi$ after a four-pulse period $T=4t_{0}$. The evolution operator in a period can be described by
\begin{equation}\label{eq:ut}
	\begin{split}
		\mathcal{U}(T)=&[R(\pi-\phi)U(t_{0})R(\phi)U(t_{0})]^{2},
	\end{split}
\end{equation}
where $U(t_{0})=e^{-iH_{L}t_{0}}$ for $\Delta^{\prime}(t)=\Delta_{0}^{\prime}$. With the relation $R(\theta_{1}+\theta_{2})=R(\theta_{1})R(\theta_{2})$, $R(\theta)aR(-\theta)=ae^{-i\theta}$, the left of Eq. (\ref{eq:ut}) can be separated into four parts as
\begin{widetext}
  \begin{gather}
    R(\pi-\phi)U(t_{0})R(\phi-\pi)=\mathrm{Exp}\{-it_{0}(-\Delta_{0}^{\prime} a^{\dag}a+\omega_{\mathrm{m}}b^{\dag}b-(Ga^{\dag}e^{-i\phi}+G^{*}ae^{i\phi})(b+b^{\dag}))\},\\
    R(\pi)U(t_{0})R(-\pi)=\mathrm{Exp}\{-it_{0}(-\Delta_{0}^{\prime} a^{\dag}a+\omega_{\mathrm{m}}b^{\dag}b-(Ga^{\dag}+G^{*}a)(b+b^{\dag}))\},\\
    R(-\phi)U(t_{0})R(\phi)=\mathrm{Exp}\{-it_{0}(-\Delta_{0}^{\prime} a^{\dag}a+\omega_{\mathrm{m}}b^{\dag}b+(Ga^{\dag}e^{-i\phi}+G^{*}ae^{i\phi})(b+b^{\dag}))\},\\
    U(t_{0})=\mathrm{Exp}\{-it_{0}(-\Delta_{0}^{\prime} a^{\dag}a+\omega_{\mathrm{m}}b^{\dag}b+(Ga^{\dag}+G^{*}a)(b+b^{\dag}))\},
  \end{gather}
\end{widetext}
where we have also used the identity $R(\pi)=R(-\pi)$, because $R(2\pi)$ is trivial in the evolution. In the approximation that the pulse interval satisfies $t_{0}\ll \omega_{m}^{-1}$, we can use the two-order Baker-Campbell-Hausdorff formula $e^{X+Y}\approx X+Y+\frac{1}{2}[X,Y]$. Then a Hamiltonian that satisfied $\mathcal{U}(T)=e^{-iH_{0}T}$ can be obtained as $H_{0}=H_{\mathrm{eff}}+H^{\prime}$, where
\begin{equation}\label{eq:heff}
	H_{\mathrm{eff}}=\omega_{\mathrm{m}}b^{\dag}b+\sigma(b+b^{\dag})^{2},
\end{equation}
\begin{equation}
	\sigma=\frac{1}{2}|G|^{2}t_{0}\mathrm{sin}\phi,
\end{equation}
and $H^{\prime}=-\Delta_{0}^{\prime} a^{\dag}a+\Delta_{0}^{\prime}(\mu a^{\dag}+\mu^{*}a)(b+b^{\dag})+\omega_{\mathrm{m}}(\mu a^{\dag}-\mu^{*}a)(b-b^{\dag})$, $\mu=\frac{1}{4}Gt_{0}(1+e^{i\phi})$. The second term in the left of Eq. (\ref{eq:heff}) is what we are interested for generating mechanical squeezing. And fortunately, the last two terms in $H^{\prime}$ can be neglected in the condition that $|G\mathrm{sin}\phi|\gg |\Delta_{0}^{\prime}|,\omega_{\mathrm{m}}$ and $|\mu|\ll1$. It means that our pulsed scheme can equivalently transpose the optomechanical coupling into a constant modulation of the potential of the mechanical oscillator. The mechanical mode will dynamically evolve into a squeezed state. For $\sigma>-\frac{\omega_{\mathrm{m}}}{4}$, Eq. (\ref{eq:heff}) can be rewritten in the classical view as
\begin{equation}
	H_{\mathrm{eff}}=\omega_{\mathrm{s}}(\sqrt{\frac{\omega_{\mathrm{m}}+4\sigma}{\omega_{\mathrm{m}}}}X_{M}^{2}+\sqrt{\frac{\omega_{\mathrm{m}}}{\omega_{\mathrm{m}}+4\sigma}}P_{M}^{2}),
\end{equation}
where $X_{M}=b^{\dag}+b$, $P_{M}=i(b^{\dag}-b)$ are mechanical quadratures and $\omega_{s}$ is the effective frequency as
\begin{equation}\label{eq:os}
	\omega_{s}=\sqrt{\omega_{m}(\omega_{m}+4\sigma)}.
\end{equation}
For $\sigma<-0.25\omega_{m}$, the mechanical mode evolves exponentially with an exponential gain as
\begin{equation}
\epsilon=\sqrt{-\omega_{m}(\omega_{m}+4\sigma)}.
\end{equation}

\section{Quantum master equations\label{ap:D}}

In order to numerically analyze the mechanical squeezing, we write down the quantum master equations as
\begin{equation}\label{eq:rho}
	\begin{split}
		\dot{\rho}=&i[\rho,H_{L}]+\frac{\kappa}{2}(2a\rho a^{\dag}-a^{\dag}a\rho-\rho a^{\dag}a)\\
		&+\frac{\gamma}{2}(n_{\mathrm{th}}+1)(2b\rho b^{\dag}-b^{\dag}b\rho-\rho b^{\dag}b)\\
		&+\frac{\gamma}{2}n_{\mathrm{th}}(2b^{\dag}\rho b-bb^{\dag}\rho-\rho bb^{\dag}).
	\end{split}
\end{equation}
It's useless and difficult to calculate the whole density matrix. We are only concerned about the evolution of quadrature variances, which can be obtained from the average values of the second-order operators, $\langle a^{\dag}a\rangle$, $\langle b^{\dag}b\rangle$, $\langle ab^{\dag}\rangle$, $\langle ab\rangle$, $\langle a^{2}\rangle$, $\langle b^{2}\rangle$. They are determined by a system of linear equations as
\begin{equation}\label{eq:oo}
	\partial_{t}\langle \hat{o}_{i}\hat{o}_{j}\rangle=Tr(\dot{\rho}\hat{o}_{i}\hat{o}_{j})=\sum_{k,l}\eta_{{k,l}}\langle \hat{o}_{k}\hat{o}_{l}\rangle,
\end{equation}
where $\hat{o}_{i,j,k}$ are operators $a$, $b$, $a^{\dag}$, $b^{\dag}$ and $\eta_{k,l}$ can be obtained from Eq. (\ref{eq:rho}). Analytical results can be obtained by replacing $H_{L}$ with $H_{\mathrm{eff}}$ in Eq. (\ref{eq:rho}).

Analytical results can be obtained by replacing $H_{L}$ with $H_{\mathrm{eff}}$ in Eq. (\ref{eq:rho}). Then the system is governed by
\begin{equation}
	\frac{d}{dt}\langle b^{\dag}b\rangle=-\gamma\langle b^{\dag}b \rangle -2i\sigma\langle b^{\dag2}\rangle+2i\sigma\langle b^{2}\rangle+\gamma n_{\mathrm{m}},
\end{equation}
\begin{equation}
	\frac{d}{dt}\langle b^{2}\rangle=\left(-\gamma-2i\omega_{\mathrm{m}}-4i\sigma\right)\langle b^{2} \rangle -2i\sigma\left(\langle b^{\dag}b\rangle+\langle bb^{\dag}\rangle\right),
\end{equation}
where $n_{\mathrm{m}}$ is the thermal occupation number of the environment, and the optical mode is neglected. For $\sigma>-\frac{\omega_{\mathrm{m}}}{4}$, the exact solution is
\begin{widetext}
\begin{equation}
	\begin{split}
		\langle b^{\dag}b\rangle(t)=&-\frac{4\sigma^{2}}{\omega_{\mathrm{s}}^{2}}\left\{a+e^{-\gamma t}\left[\left(n_{\mathrm{th}}-a\right)\mathrm{cos}(2\omega_{\mathrm{s}}t)+\frac{\gamma}{2\omega_{\mathrm{s}}}\left(n_{\mathrm{m}}-a\right)\mathrm{sin}(2\omega_{\mathrm{s}}t)\right]\right\}\\
		&+\frac{(\omega_{\mathrm{m}}+2\sigma)^{2}}{\omega_{\mathrm{s}}^{2}}\left[n_{\mathrm{m}}+(n_{\mathrm{th}}-n_{\mathrm{m}})e^{-\gamma t}\right],
	\end{split}
\end{equation}
\begin{equation}
	\begin{split}
		\mathrm{Re}\langle b^{2}\rangle(t)&=\frac{2\sigma(\omega_{\mathrm{m}}+2\sigma)}{\omega_{\mathrm{s}}^{2}}\\
		&\left\{a-n_{\mathrm{m}}+e^{-\gamma t}\left[\left(n_{\mathrm{th}}-a\right)\mathrm{cos}(2\omega_{\mathrm{s}}t)+\frac{\gamma}{2\omega_{\mathrm{s}}}\left(n_{\mathrm{m}}-a\right)\mathrm{sin}(2\omega_{\mathrm{s}}t)+n_{\mathrm{m}}-n_{\mathrm{th}}\right]\right\}
	\end{split},
\end{equation}
\begin{equation}
	\mathrm{Im}\langle b^{2}\rangle(t)=-\frac{2\sigma}{\omega_{\mathrm{s}}}\left\{b\left[1-e^{-\gamma t}\left(\mathrm{cos}(2\omega_{\mathrm{s}}t)+\frac{\gamma}{2\omega_{\mathrm{s}}}\mathrm{sin}(2\omega_{\mathrm{s}}t)\right)\right]+e^{-\gamma t}(n_{\mathrm{th}}+\frac{1}{2})\mathrm{sin}(2\omega_{\mathrm{s}}t)\right\},
\end{equation}
with $n_{\mathrm{th}}$ the initial phonon number and
\begin{equation}
	a=\mathrm{Re}\left(\frac{\gamma n_{\mathrm{m}}+i\omega_{\mathrm{s}}}{\gamma-2i\omega_{\mathrm{s}}}\right),b=\mathrm{Im}\left(\frac{\gamma n_{\mathrm{m}}+i\omega_{\mathrm{s}}}{\gamma-2i\omega_{\mathrm{s}}}\right).
\end{equation}
The quadrature variances satisfy $\delta X_{M}^{2}=1+2(\langle b^{\dag}b\rangle+\langle b^{2}\rangle)$ and $\delta P_{M}^{2}=1+2(\langle b^{\dag}b\rangle-\langle b^{2}\rangle)$. The variance of squeezed quadrature $Y_{M}$ satisfies $\delta Y_{M}^{2}=1+2(\langle b^{\dag}b\rangle-|\langle b^{2}\rangle|)$. In the short-time approximation $\omega_{\mathrm{m}}t_{0}\ll1$ and $\gamma n_{\mathrm{m}}\ll1$, the evolution of quadrature variances can be obtained as
\begin{equation}
	\delta P_{M}^{2}=(1+2n_{\mathrm{th}})\left[1+\frac{2\sigma}{\omega_{\mathrm{m}}}(1-\mathrm{cos}2\omega_{\mathrm{s}}t)\right],
\end{equation}
\begin{equation}
	\delta X_{M}^{2}=(1+2n_{\mathrm{th}})\left[1-\frac{2\sigma}{\omega_{\mathrm{m}}+4\sigma}(1-\mathrm{cos}2\omega_{\mathrm{s}}t)\right],
\end{equation}
\begin{equation}
	\delta Y_{M}^{2}=1+2\left\{n_{\mathrm{th}}+\frac{2}{\omega_{\mathrm{s}}^{2}}(2n_{\mathrm{th}}+1)\mathrm{sin}^{2}(\omega_{\mathrm{s}}t)\left[2\sigma^{2}-|\sigma|\sqrt{(\omega_{\mathrm{m}}+2\sigma)^{2}+\omega_{\mathrm{s}}^{2}\mathrm{cot}^{2}(\omega_{\mathrm{s}}t)}\right]\right\}.
\end{equation}
Then the squeezing limit is given by (corresponding to Eq. (9) in the main text)
\begin{equation}\label{eq:rm}
	(\delta Y_{M}^{2})_{\mathrm{min}}=\left\{
	\begin{aligned}
		&(1+2n_{\mathrm{th}})\frac{\omega_{\mathrm{m}}+4\sigma}{\omega_{\mathrm{m}}},&-\frac{\omega_{\mathrm{m}}}{4}<\sigma<0\\
		&(1+2n_{\mathrm{th}})\frac{\omega_{\mathrm{m}}}{\omega_{\mathrm{m}}+4\sigma},&\sigma>0~~~~~
	\end{aligned}
	\right.
\end{equation}
which is approached at $\omega_{\mathrm{s}}t=n\pi+\pi/2$, $n=0,1,2,\cdots$. There is maximal squeezing degree at the first point of time $t_{\mathrm{m}}=\frac{\pi}{2\omega_{\mathrm{s}}}$, if considering the influence of the decay rates. In the maximally squeezed state, quadrature $P_{M}$ is squeezed for $-\frac{\omega_{\mathrm{m}}}{4}<\sigma<0$ while $X_{M}$ is squeezed for $\sigma>0$.

By contrast in the long-time approximation, the steady value of quadrature variances reads
\begin{equation}
	\delta P_{M}^{2}=(1+2n_{\mathrm{m}})\left(1+\frac{2\sigma(\omega_{\mathrm{m}}+4\sigma)}{\gamma^{2}/4+\omega_{\mathrm{s}}^{2}}\right),
\end{equation}
\begin{equation}\label{eq:rs}
	\delta X_{M}^{2}=(1+2n_{\mathrm{m}})\left(1-\frac{2\sigma\omega_{\mathrm{m}}}{\gamma^{2}/4+\omega_{\mathrm{s}}^{2}}\right).
\end{equation}
The squeezing degree can't suppress the $3\mathrm{dB}$ limit in the steady state.

And for $\sigma=-\frac{\omega_{\mathrm{m}}}{4}$, the exact solution is
\begin{equation}
  \langle b^{\dag}b\rangle(t)=\frac{\omega_{\mathrm{m}}^{2}}{2\gamma^{2}}\left[\gamma^{2} t^{2}(n_{\mathrm{th}}-n_{\mathrm{m}})-(\gamma t+1)(2n_{\mathrm{m}}+1)\right]e^{-\gamma t}+(n_{\mathrm{th}}-n_{\mathrm{m}})e^{-\gamma t}+\frac{\omega_{\mathrm{m}}^{2}}{2\gamma^{2}}(2n_{\mathrm{m}}+1)+n_{\mathrm{m}},
\end{equation}
\begin{equation}
  \mathrm{Re}\langle b^{2}\rangle(t)=\frac{\omega_{\mathrm{m}}^{2}}{2\gamma^{2}}\left[\gamma^{2} t^{2}(n_{\mathrm{th}}-n_{\mathrm{m}})-(\gamma t+1)(2n_{\mathrm{m}}+1)\right]e^{-\gamma t}+\frac{\omega_{\mathrm{m}}^{2}}{2\gamma^{2}}(2n_{\mathrm{m}}+1),
\end{equation}
\begin{equation}
  \mathrm{Im}\langle b^{2}\rangle(t)=\frac{\omega_{\mathrm{m}}}{2\gamma}\left[2\gamma t(n_{\mathrm{th}}-n_{\mathrm{m}})-(2n_{\mathrm{m}}+1)\right]e^{-\gamma t}+\frac{\omega_{\mathrm{m}}}{2\gamma}(2n_{\mathrm{m}}+1).
\end{equation}
Letting $\gamma=0$, we obtained
\begin{equation}
	\delta Y_{M}^{2}(t)=(1+2n_{\mathrm{th}})\left[1-\frac{\omega_{\mathrm{m}}^{2}t^{2}}{2}\left(\sqrt{1+\frac{4}{\omega_{\mathrm{m}}^{2}t^{2}}}-1\right)\right].
\end{equation}

For $\sigma<-\frac{\omega_{\mathrm{m}}}{4}$ and $\gamma\ne\epsilon$, the exact solution is
\begin{equation}
  \begin{split}
    \langle b^{\dag}b\rangle(t)=&\left[-\omega_{0}^{2}+8\sigma^{2}(e^{2\epsilon t}+e^{-2\epsilon t})\right]\frac{n_{\mathrm{th}}}{4\epsilon^{2}}e^{-\gamma t}\\
    &+\frac{\omega_{0}^{2}+\gamma^{2}}{\gamma^{2}-4\epsilon^{2}}n_{\mathrm{m}}-\left[-\omega_{0}^{2}+8\sigma^{2}\left(\frac{\gamma}{\gamma-2\epsilon}e^{2\epsilon t}+\frac{\gamma}{\gamma+2\epsilon}e^{-2\epsilon t}\right)\right]\frac{n_{\mathrm{m}}}{4\epsilon^{2}}e^{-\gamma t}\\
    &+\frac{8\sigma^{2}}{\gamma^{2}-4\epsilon^{2}}-\left[\frac{1}{\gamma-2\epsilon}e^{2\epsilon t}-\frac{1}{\gamma+2\epsilon}e^{-2\epsilon t}\right]\frac{2\sigma^{2}}{\epsilon}e^{-\gamma t},
  \end{split}
\end{equation}
\begin{equation}
  \begin{split}
    \mathrm{Re}\langle b^{2}\rangle(t)=&\left[1-\frac{1}{2}(e^{2\epsilon t}+e^{-2\epsilon t})\right]\frac{\sigma\omega_{0}n_{\mathrm{th}}}{\epsilon^{2}}e^{-\gamma t}\\
    &-\frac{4\sigma\omega_{0}}{\gamma^{2}-4\epsilon^{2}}n_{\mathrm{m}}-\left[1-\left(\frac{\gamma}{\gamma-2\epsilon}e^{2\epsilon t}+\frac{\gamma}{\gamma+2\epsilon}e^{-2\epsilon t}\right)\right]\frac{\sigma\omega_{0}n_{\mathrm{m}}}{\epsilon^{2}}e^{-\gamma t}\\
    &-\frac{2\sigma\omega_{0}}{\gamma^{2}-4\epsilon^{2}}+\left[\frac{1}{\gamma-2\epsilon}e^{2\epsilon t}-\frac{1}{\gamma+2\epsilon}e^{-2\epsilon t}\right]\frac{\omega_{0}\sigma}{2\epsilon}e^{-\gamma t},
  \end{split}
\end{equation}
\begin{equation}
  \begin{split}
    \mathrm{Im}\langle b^{2}\rangle(t)=&-(e^{2\epsilon t}-e^{-2\epsilon t})\frac{\sigma n_{\mathrm{th}}}{\epsilon}e^{-\gamma t}\\
    &-\frac{4\sigma\gamma}{\gamma^{2}-4\epsilon^{2}}n_{\mathrm{m}}+\left(\frac{1}{\gamma-2\epsilon}e^{2\epsilon t}-\frac{1}{\gamma+2\epsilon}e^{-2\epsilon t}\right)\frac{\sigma \gamma n_{\mathrm{m}}}{\epsilon}e^{-\gamma t}\\
    &-\frac{2\gamma\sigma}{\gamma^{2}-4\epsilon^{2}}+\left[\frac{1}{\gamma-2\epsilon}e^{2\epsilon t}+\frac{1}{\gamma+2\epsilon}e^{-2\epsilon t}\right]\sigma e^{-\gamma t}.
  \end{split}
\end{equation}
Letting $\gamma=0$, we obtained
\begin{equation}
	\delta Y_{M}^{2}(t)=(1+2n_{\mathrm{th}})\left\{1-\frac{2\sigma^{2}}{\epsilon^{2}}(1-e^{-2\epsilon t})\left[\sqrt{(e^{2\epsilon t}-1)^{2}+e^{2\epsilon t}\frac{\epsilon^{2}}{\sigma^{2}}}-(e^{2\epsilon t}-1)\right]\right\}.
\end{equation}
\end{widetext}

\section{Wigner Function\label{ap:E}}

In this section, we will provide the detailed calculation of the Wigner function where we can see the squeezing dynamics and performance visually. The initial states of optical and mechanical modes are both thermal states, the Wigner functions of which can be written as
\begin{equation}
	W(q,p)=\frac{1}{\pi(\langle n\rangle+1/2)}e^{-(q^{2}+p^{2})/(\langle n\rangle+1/2)},
\end{equation}
where $\langle n\rangle$ is the average occupation number. In the optomechanical system, it's more useful to calculate the Wigner function in total phase space including both optical and mechanical quadratures, which is
\begin{equation}\label{eq:total}
	\begin{split}
		W_{\mathrm{total}}&(X_{L},P_{L},X_{M},P_{M},t)=\\
		&W_{L}(X_{L},P_{L},t)\times W_{M}(X_{M},P_{M},t).
	\end{split}
	\end{equation}
Here $W_{L}(X_{L},P_{L},t)$ and $W_{M}(X_{M},P_{M},t)$ are the Wigner functions of the optical and mechanical modes, which can be obtained from the integral of Eq. (\ref{eq:total}) over the other phase space. The Wigner function of the initial state is
\begin{equation}
	W_{\mathrm{total}}(\bm{X},0)=\left(\frac{1}{\pi(n_{\mathrm{th}}+1/2)}\right)^{2}e^{-\bm{X}^{T}\bm{X}/(n_{\mathrm{th}}+1/2)},
\end{equation}
where we have defined $\bm{X}=(X_{L},P_{L},X_{M},P_{M})^{T}$ and $n_{\mathrm{th}}$ is the initial phonon number. The evolution of quadrature "vector" $\bm{X}$ is described by
\begin{equation}
	\bm{X}(t)=\mathcal{U}(t)\bm{X}(0)\equiv \mathcal{A}\bm{X}(0),
\end{equation}
where $\mathcal{U}(t)$ is the evolution operator and $\mathcal{A}$ is a matrix that denotes the linear evolution of $\bm{X}(t)$. Then the time evolution of the total Wigner function can be obtained as
\begin{equation}
	W_{\mathrm{total}}(\bm{X},t)=\left(\frac{1}{\pi(n_{\mathrm{th}}+1/2)}\right)^{2}e^{-\left(\mathcal{A}\bm{X}\right)^{T}\mathcal{A}\bm{X}/(n_{\mathrm{th}}+1/2)}.
\end{equation}
And the optical and mechanical Wigner functions are given by
\begin{equation}
	W_{L}(X_{L},P_{L},t)=\int_{-\infty}^{\infty}dX_{M}\int_{-\infty}^{\infty}dP_{M} W_{\mathrm{total}}(\bm{X},t),
\end{equation}
\begin{equation}
	W_{M}(X_{M},P_{M},t)=\int_{-\infty}^{\infty}dX_{L}\int_{-\infty}^{\infty}dP_{L}W_{\mathrm{total}}(\bm{X},t).
\end{equation}

\begin{figure}[t]
	\includegraphics[width=0.48\textwidth]{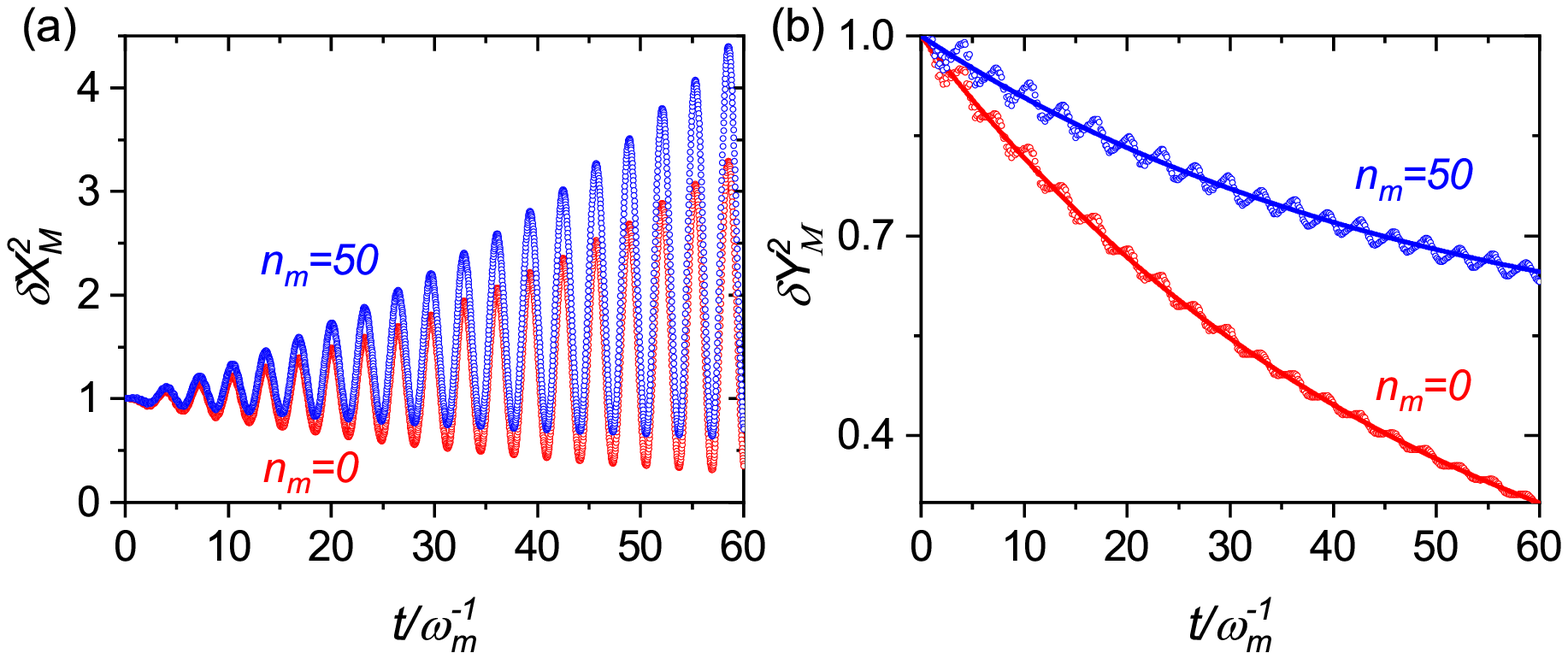}%
	\caption{(a),(b) The evolution of variances of mechanical quadratures $X_{M}$ (a) and $Y_{M}$ (b) for $n_{\mathrm{m}}=0$ in red and $n_{\mathrm{m}}=50$ in blue. In panel (b), the theoretical evolution (solid line) given by Eq. (\ref{eq:pa}) agrees well with the numerical results (open circles) based on the driving strength modulated as Eq. (\ref{eq:omgt}).\label{fig:all}}
\end{figure}

\section{Parametric resonance\label{ap:F}}

Another efficient way to generate mechanical squeezing in the optomechanical system is to consider the mechanical mode as a parametric oscillator, which can also be used in our model. Parametric resonance occurs when the external driving strength is modulated as
\begin{equation}\label{eq:omgt}
	\Omega(t)=\Omega_{0}\mathrm{sin}(\omega_{s}t),
\end{equation}
with $\Omega_{0}$ being a constant and $\omega_{\mathrm{s}}$ given by Eq. (\ref{eq:os}) where we find numerically $G\approx 0.45g\Omega_{0}t_{0}$. Noted that the average values $\alpha$ and $\beta$ are time-dependent but can also be calculated by master equations. In Fig. \ref{fig:all}, we plot the evolution of mechanical quadrature variances in the parametric resonance case. The squeezing performance is limited by the thermal occupation number $n_{\mathrm{m}}$. The numerical results agree well with the adiabatic theory given by Jieqiao Liao {\it et al} \cite{PhysRevA.83.033820}. Starting from the vacuum state, the evolution of squeezed quadrature can be described by
\begin{equation}\label{eq:pa}
	\delta Y_{M}^{2}(t)\approx e^{-(\gamma+\xi_{0})t}+\frac{\gamma(2n_{m}+1)}{\gamma+\xi_{0}}(1-e^{-(\gamma+\xi_{0})t}),
\end{equation}
where $\xi_{0}=|\omega_{s}-\omega_{m}|$ describes the parametric gain in parametric resonance. Unlike the large-detuning system in Ref. \cite{PhysRevA.83.033820}, $\xi_{0}$ can be arbitrarily large with appropriate parameters, i.e. $t_{0}\to0$, $G\to\infty$ and $\phi=\pi/2$, which also has potential in the generation of strong mechanical squeezing in the steady state \cite{PhysRevLett.107.213603}.

\section{Discussion about the detuning-switched driving}
\label{ap:dis}

To obtain the mechanical squeezing with detuning-switched driving, an important requirement is that the cavity must oscillate following the laser driving, i.e. the cavity must react to the large detuning. We discuss this requirement from two different aspects.

First, we consider a general classical Langevin equation describing a single-mode cavity
\begin{equation}
  \dot{a}(t)=(i\omega_{0}-\kappa/2)a(t)+F(t),
\end{equation}
where $a(t)$ is the classical cavity field, $\omega_{0}$ is the cavity resonance frequency, $\kappa$ is the cavity decay rate and $F(t)=Ae^{i\omega t}$ is a harmonic drive. The solution after applying the drive ($t>t_{0}$) is given by
\begin{equation}
  \begin{split}
    a(t)&=e^{(i\omega_{0}-\kappa/2)(t-t_{0})}[a(t_{0})-\frac{Ae^{i \omega t_{0}}}{i(\omega-\omega_{0})+\kappa/2}]\\
    &+e^{i\omega (t-t_{0})}\frac{Ae^{i \omega t_{0}}}{i(\omega-\omega_{0})+\kappa/2}.
  \end{split}
\end{equation}
Initially the cavity is driven by a laser $F_{1}(t)=A_{1}e^{i\omega_{1}t}$ near the resonance $\omega_{1}\approx\omega_{0}$, the cavity field will reach a steady-state amplitude $\left|\frac{A_{1}}{i(\omega_{1}-\omega_{0})-\kappa/2}\right|$. Assume at $t=0$ we suddenly increase the detuning to a large value with a driving $F_{2}=A_{2}e^{i\omega_{2}t}$ and an initial condition $a(0)=\frac{A_{1}}{i(\omega_{1}-\omega_{0})+\kappa/2}$, the following evolution will be
\begin{equation}
  \begin{split}
    a(t)=&e^{(i\omega_{0}-\kappa/2)t}[\frac{A_{1}}{i(\omega_{1}-\omega_{0})+\kappa/2}-\frac{A_{2}}{i(\omega_{2}-\omega_{0})+\kappa/2}]\\
    +&e^{i\omega_{2} t}\frac{A_{2}}{i(\omega_{2}-\omega_{0})+\kappa/2}.
  \end{split}
\end{equation}
Consequently, to make the cavity field change from frequency $\omega_{1}$ to $\omega_{2}$ and acquire the rotating phase we wanted, the driving laser should satisfy
\begin{equation}
  \frac{A_{1}}{i(\omega_{1}-\omega_{0})+\kappa/2}=\frac{A_{2}}{i(\omega_{2}-\omega_{0})+\kappa/2}.
\end{equation}
It means that when we increase the detuning of the laser, we also need to increase the power of the laser accordingly. Then in the frame with laser frequency, although the optomechanical coupling strength remains constant, the cavity field will rotate and acquire a phase; and in the frame with cavity resonance frequency, the cavity field does not rotate but the optomechanical coupling strength will rotate, leading to a new kind of squeezing effect in the mechanical oscillator.

Second, an optical cavity differs from a cavity described by the single-mode Langevin equation. An optical cavity is not a point object and can not react instantaneously to the sudden change of a driving laser, as everything propagates at the speed of light. This propagation effect is not captured by the Langevin equation, which only provides the amplitude-phase degrees of freedom.

The finite propagation speed of the light can be captured if multiple azimuthal modes are involved. The
frequency domain picture of this system is periodic Lorentzian response functions spaced by the FSR. The inverted time domain response is a periodic near-delta function spaced by the round-trip time, which captures the round-trip dynamics of the input pulse bouncing back and forth between the two mirrors. Therefore, when there is a sudden change of the input field, instead of immediately forming a new oscillation frequency inside the cavity, it could well be that this sudden change results in some localized pulse structure in space and time, and bounce back and forth inside the cavity for a very long time until the cavity decays to a new equilibrium state. It is far away from what we want in the squeezing mechanism, where we want an immediate switch of the oscillation frequency. Therefore the cavity field needs to switch to a new equilibrium instantly.

The frequency domain picture provides some insights on how to avoid such a scenario. When there is a sudden switch of the input laser, this step-like change contains extremely broad frequency components, and can excite multiple optical modes spanning many FSRs. Such an excitation in the time domain is the resulting localized intracavity structure that requires a very long time to reach a new equilibrium. Therefore, the laser frequency (also amplitude) switching should not be arbitrarily fast in the sense that other azimuthal optical modes should not be excited. The switching needs to be slower than the FSR, or the round trip time of the cavity, so the cavity field can adiabatically reach a new equilibrium during the switching. Therefore, the requirement on the laser switching scheme is that it should not be faster than the round trip time, but should be a lot faster than all the other time scales in the experiment, e.g. the inverse of $G$, $\kappa$, $\omega_{\mathrm{m}}$, etc.

Realistically, this issue might not pose any problem for optical implementations, since the FSR in optical domains is usually $\mathrm{GHz}$ range for well-established platforms. However, it could be a realistic concern for microwave platforms where switching of the microwave fields can be really fast, faster than the microwave frequency, and the frequency spacing to higher order modes (FSR).

\section{Experimental realization}

The optomechanical system we considered is in the strong coupling regime ($G>\omega_{\mathrm{m}}$), and the squeezing mechanism requires the optical cavity can respond quickly to the detuning of the driving laser, which limits that the cavity decay rate can not be too small. Such a condition is possible with an experimental setup of a three-dimensional microwave cavity \cite{peterson_ultrastrong_2019}. The mechanical frequency is around $\omega_{\mathrm{m}}\approx 9.696\mathrm{MHz}$, the single-photon optomechanical coupling strength is $g\approx167\mathrm{Hz}$, and the cavity decay rate is $\kappa\approx1  \mathrm{MHz}$. To enter the strong coupling regime, the average intra-cavity photon number should exceed $(\omega_{\mathrm{m}}/g)^{2}\approx3.37\times10^{9}$. For a detuned cavity, the average intra-cavity photon number is given by
\begin{equation}
  n_{\mathrm{cav}}=\frac{P}{\hbar \omega_{\mathrm{L}}}\frac{\kappa}{(\kappa/2)^{2}+\Delta^{2}},
\end{equation}
where $P$ is the power of the driving laser. In our model, the detuning is switched to a large value periodically. So the intra-cavity photon number is limited by the detunings instead of the cavity decay rate. For typical laser detuning $\Delta=10\omega_{\mathrm{m}}$, the laser power should exceed $136\mathrm{\mu W}$ (at frequency $6.5\mathrm{GHz}$). Consequently, the squeezing mechanism proposed here is realizable with existing platforms, and the mechanical squeezing can be observed through an additional probe beam.


\end{document}